# Hybrid stochastic-mechanical modeling of precipitation thresholds of shallow landslide initiation

Edoardo Rundeddu[1], José J. Lizárraga[2], Giuseppe Buscarnera[2][3]

**Abstract**

Numerous early warning systems based on rainfall measurements have been designed over the last decades to forecast the onset of rainfall-induced shallow landslides. However, their use over large areas poses challenges due to uncertainties related with the interaction among various controlling factors. We propose a hybrid stochastic-mechanical approach to quantify the role of the hydro-mechanical factors influencing slope stability and rank their importance. The proposed methodology relies on a physically-based model of landslide triggering, and a stochastic approach treating selected model parameters as correlated aleatory variables. The features of the methodology are illustrated by referencing data for Campania, an Italian region characterized by landslide-prone volcanic deposits. Synthetic intensity-duration (ID) thresholds are computed through Monte Carlo simulations. Several key variables are treated as aleatoric, constraining their statistical properties through available measurements. The variabilities of topographic features (e.g., slope angle), physical and hydrological properties (e.g., porosity, dry unit weight $\gamma_d$, and saturated hydraulic conductivity, $K_s$), and pre-rainstorm suction is evaluated to inspect its role on the resulting scatter of ID thresholds. We find that: i) $K_s$ is most significant for high-intensity storms; ii) in steep slopes, changes in pressure head greatly reduce the timescale of landslide triggering, making the system heavily reliant on initial conditions; iii) for events occurring at long failure times (gentle slopes and/or low intensity storms), the significance of the evolving stress level (through $\gamma_d$) is highest. The proposed approach can be translated to other regions, expanded to encompass new aleatory variables, and combined with other hydro-mechanical triggering models.

## 1. Introduction

The expansion of densely populated urban agglomerates and the increasing frequency of extreme weather events taking place during the last few decades is greatly exacerbating the risks for life and property caused by landslides (Haque et al., 2019; Schuster and Highland, 2001). The United States alone has been reported to suffer yearly losses of over $1 billion due to damage to public and private property (Fleming and Taylor, 1980); the average yearly cost of geohydrological hazards in Italy between 1944 and 2012 was €0.9 billion (Donnini et al., 2017). This has motivated the rise in interest in Landslides Early Warning Systems (LEWS) (Guzzetti et al., 2020), i.e., technologies designed to identify precursors of fast, life-threatening landslides through precipitation intensity-duration thresholds (Caine, 1980). Currently, various LEWS systems around the world use historical rainfall data to express such intensity-duration thresholds in probabilistic terms (see examples from Italy, Washington state and Korea; Aleotti (2004), Scheevel et al. (2017), Park et al. (2019)). While historical databases provide empirical evidence of intensity-duration thresholds, such data is

[1] Department of Mechanical Engineering, Northwestern University, Evanston (IL), USA
[2] Department of Civil and Environmental Engineering, Northwestern University, Evanston (IL), USA
[3] Corresponding author. E-mail address: g-buscarnera@northwestern.edu

characterized by a large scatter in failure times when landslide events from large, and especially topographically complex regions are grouped together (Guzzetti et al., 2007).

Decoding how topography, soil properties and initial hydrologic conditions affect the reported failure times exclusively from historical datasets is therefore a remarkable challenge. In this context, triggering models based on the principles of hydrology and mechanics of soils such as TRIGRS (Baum et al., 2002) and SHALSTAB (Dietrich and Montgomery, 1998) are convenient tools to guide the design of physics-based LEWS. The benefits of these methods emerge especially when used in conjunction with computing-visualization platforms based on Geographic Information Systems (GIS), through which it is possible to account explicitly for the regional variability of topographic and hydro-mechanical conditions (i.e., two factors greatly augmenting the data scatter of empirical intensity-duration thresholds). Such arguments are corroborated by recent results reported by Fusco et al. (2019), who examined the sensitivity of intensity-duration thresholds to the slope angle with infiltration analyses. Their results showed that, while slope angle plays a minor role if compared to the large variability of the intensity-duration thresholds proposed in the literature, at lower intensities changes in slope angle of 5° account for tens of hours of difference in failure times in the location studied. It is therefore arguable that accurate measurements of the slope angle mapped across a region of interest should be used to optimize LEWS thresholds. In fact, from a LEWS user's perspective, slope angle data for the location of interest is indeed likely to be as accessible as intensity and real-time precipitation duration data: digital elevation models (DEM) can now be obtained with unprecedented accuracy using methods such as airborne Lidar techniques (Roering et al., 2013). However, databases that report slope angle data of historical landslides are not as common, and the literature often groups landslide events associated with different slope angles on the same intensity-duration plot.

In this paper, we propose a hybrid strategy aimed at satisfying two competing needs: (i) accounting for measurable variables which play a distinct role in the hydro-mechanics of landslides and can be ascertained in near-deterministic form (e.g., slope angle); (ii) encompassing the aleatory nature of controlling factors, which, although measurable in principle, can be known only within a range of statistical variability (e.g., seasonality of initial conditions, uncertain hydro-mechanical properties). For this purpose, we pursue a stochastic modeling strategy based on Monte Carlo simulations, which allow for the characterization of soil properties exhibiting natural variability or characterized by significant uncertainty (Liu, 2007; Lee et al., 2013; Peres and Cancelliere, 2014) and that has been used with considerable success in the domain of landslide triggering analyses (Weidner et al., 2018; Lizarraga and Buscarnera, 2019). Specifically, here we use physically-based Monte Carlo analyses to generate synthetic intensity-duration thresholds embedding the role of the slope angle. The goal is to provide a platform to explain and reduce the source of large data scatter in intensity-duration thresholds for LEWS applications. The Monte Carlo variables include mechanical and hydraulic soil properties, as well as initial pressure head. The analyses are carried out through a C++/MATLAB solver simulating water infiltration in one-dimensional soil columns, while accounting for the effect of the correlation between hydro-mechanical variables on the margins of safety. A strategy to evaluate the importance of each Monte Carlo variable across different slope angle and rainfall intensity regimes is also proposed; this methodology aims at narrowing the probabilistic variability of intensity-duration thresholds by recommending which aleatory variables should be further studied depending on the topographic and meteorological characteristics of a region of interest. To better illustrate the key features of the proposed methodology, the model is tested against data published in the literature for an extensively studied case study in Campania (Italy) for which baseline estimates of the governing hydro-mechanical properties have been previously calibrated (Lizarraga et al., 2017).

## 2. Methodology

*2.1 – Water infiltration model and safety factor determination*

The water infiltration model used in this paper relies on Richards equation for water mass balance in a one-dimensional soil column (Richards, 1931):

$$n_p C_w(h) \frac{\partial h}{\partial t} = \frac{\partial}{\partial z}\left[K(h)\left(\frac{\partial h}{\partial z} + 1\right)\right] \quad (1)$$

where $n_p$ is the soil porosity; $h$ is the pressure head; $C_w(h)$ is the unsaturated storage coefficient function; $K(h)$ is the hydraulic conductivity function; $z$ is the depth from the surface. The literature has proposed numerous models for $C_w(h)$ and $K(h)$ (eg. Van Genuchten, 1980; Campbell, 1974). This paper utilizes an exponential-form relation (Gardner, 1958) to characterize the hydraulic conductivity function and the water retention curve to minimize the number of parameters not amenable to Monte Carlo variables. Additionally, this model has already been used and calibrated for the selected case study (Lizarraga and Buscarnera, 2019). The relevant functions, including calculation of $S_r$ (degree of saturation) are reported below:

$$S_r(h) = \begin{cases} \frac{1}{n_p}(\theta_r + (n_p - \theta_r)e^{bh}) & h < 0 \\ 1 & h \geq 0 \end{cases} \quad (3)$$

$$C_w(h) = b(n_p - \theta_r)e^{bh} \quad (4)$$

$$K(h) = \begin{cases} K_s e^{bh} & h < 0 \\ K_s & h \geq 0 \end{cases} \quad (5)$$

where $\theta_r$ is the residual volumetric water content; $K_s$ is the saturated hydraulic conductivity; $b$ is the model fitting parameter. By convention, this paper considers $h < 0$ to be the unsaturated soil domain, whereas $h > 0$ corresponds to saturation.

The infiltration analysis is coupled with a factor of safety ($FS$) for slope stability (Lizarraga et al., 2017):

$$FS(z,t) = \frac{\tan \phi'}{\tan \alpha}\left(1 - \frac{\gamma_w h(z,t) S_r(h)}{\sigma^{net}(z,t)}\right) \quad (6)$$

where $\phi'$ is the internal friction angle of the soil; $\alpha$ is the slope angle; $\gamma_w$ is the specific weight of water; $\sigma^{net}(z)$ is the net stress at depth $z$. The inclusion of $S_r(h)$ to model suction-dependence strength defines this behavior in terms of a varying physical property instead of a calibrated constant (Nuth and Laloui, 2007), which is more representative of the different data points included in a Monte Carlo simulation. $\sigma^{net}(z,t)$ can be expressed in terms of $\gamma_d$, the soil specific dry weight:

$$\sigma^{net}(z,t) = z\gamma_d + n_p \gamma_w \int_0^z S_r(z,t)\, dz \quad (7)$$

The critical condition for instability occurs when $FS(z,t)$ first drops below 1 at any $z$, defining the corresponding failure time as $t_f$.

*2.2 – Computational discretization*

A first-order finite difference method was implemented to numerically solve Eq.1. The discretized model setup is illustrated in Fig.1.

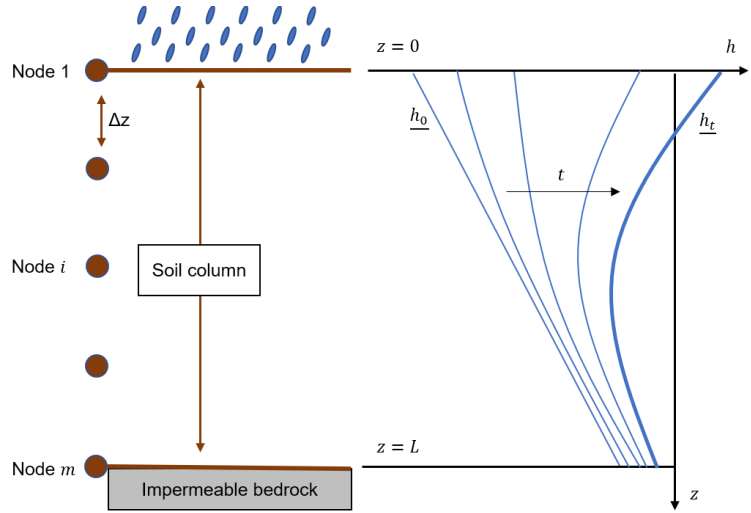

**Fig.1:** Discretized 1D soil column with top water influx and impermeable bedrock (left); the evolution of pressure head profiles starting from a linear profile $h_0$ (right).

Eq.3-5 are evaluated at every node, for a total of $m$ nodes, equally spaced by a length $\Delta z = \frac{L}{m-1}$ where $L$ is the total height of the soil column. The boundary conditions used are:

1. Constant flux of $I \cos \alpha$ m/s into node 1, where $I$ represents the magnitude of the meteorological rainfall intensity and the $\cos \alpha$ term is used to retain the flux normal to the slope.
2. Constant flux of 0 m/s out of node $m$, modelling an impermeable bedrock.

The program computes the time evolution of pressure head profile $\underline{h}$ constituted by the pressure head value $h$ at any node $i$ of the soil column. The initial conditions are based on a Monte Carlo initial pressure head profile $\underline{h_0}$, addressed in section 2.3. The factor of safety expression and net stress (Eq.6-7) are similarly discretized for each node. Further details about the model implementation are provided in Appendix A.

*2.3 – Monte Carlo variables*

The goal of variable randomization is to account for both measurement uncertainty and spatial variability of hydro-mechanical properties and initial conditions. In this paper, all Monte Carlo randomizations are based on readily measurable data from site instrumentation. For each variable, a probability density function (pdf) can be selected and fitted to the available data. Lastly, correlation between different variables is used.

The literature proposes several models to explain observed relationships between soil physical properties. Relationships between $\phi'$ and $\gamma_d$ (Picarelli et al., 2006; Bardet et al., 2011) and between $K_s$ and $n_p$ (Franzmeier, 1991; Fallico et al., 2010) have been reported. Further correlations between the variables discussed in this paper can also be found for certain locations or soil types. For the scope of this paper, to illustrate that the method proposed can support inter-variable correlation models, only the well-established relationship between $n_p$ and $\gamma_d$ is included in the Monte Carlo setup. Future applications of this methodology should be combined with a location-specific study of the soil properties to enrich the model with other potential correlations.

Focusing on this case study, Campanian soils, such as those found in the areas of Sarno and Cervinara, are typically characterized by a layer of volcanic ashes (Olivares et al., 2019). For simplicity, the soil column shown in Fig.1 is assumed to be a uniform pyroclastic layer where all nodes share the same pdf for each physical nodal property. Further improvements or applications of this model could involve defining different layers to reflect a typical geomorphological structure encountered in a site of interest by assigning different pdfs to different nodes for the same variable.

Table 1 illustrates, for each physical property, the typical range reported in the literature for the Campania region. Most of the variables ($\phi'$, $\gamma_d$, $n_p$, $L$) are characterized by relatively low variances that keep their distribution within the same order of magnitude. For each variable, a pdf was generated from a normal distribution with mean $\mu$ at the center of the variable's typical range and standard deviation $\sigma$ such that the typical range involves a 95% confidence interval. Normal distributions for some of these parameters have been used in the literature (Lee et al., 2013). Other pdfs, such as uniform distribution, have also been reported (Gorsevski et al., 2006). The normal pdf, for any random variable $X = x$, is given by Eq.8:

$$f(x) = \frac{1}{\sigma\sqrt{2\pi}} e^{-\frac{(x-\mu)^2}{2\sigma^2}} \qquad (8)$$

The typical range of $K_s$ spans various orders of magnitude and varies greatly for different types of soil (Bear, 1972). There is agreement that the lognormal distribution is a reliable pdf for modeling the stochasticity of $K_s$ (Zhai and Benson, 2005; Kosugi, 1996). The 2-parameter log-normal pdf is given by Eq.9. Note that the parameters $\mu$ and $\sigma$ do not describe the mean and standard deviation of the sample to be fitted, as is the case with the normal distribution.

$$f(x) = \frac{1}{x\sigma\sqrt{2\pi}} e^{-\frac{(\ln x - \mu)^2}{2\sigma^2}} \qquad (9)$$

While Zhai and Benson (2005) suggest that adding a third parameter to the distribution can further improve the accuracy of the pdf for hydraulic conductivity, procedures based on a standard 2-parameter log-normal pdf have also led to satisfactory results (Pirone et al., 2016) and will therefore be used here for the sake of simplicity, but without loss of generality. Finally, Eq.10 was used to obtain the log-normal pdf parameters $\mu$ and $\sigma$ from the available statistics.

**Table 1:** typical ranges and main statistics for the physical properties of Campanian soils.

| Variable | Units | Typical range | μ | σ | Distribution | Reference |
|---|---|---|---|---|---|---|
| $\phi'$ (internal friction angle) | ° | [37.0, 39.0] | 38.0 | 0.5 | Normal | Damiano and Olivares (2010) |
| $\gamma_d$ (specific soil dry weight) | kN m$^{-3}$ | [7.44, 12.62] | 10.03 | 1.29 | Normal | Pirone et al. (2016) |
| $n_p$ (porosity) | - | [0.68, 0.75] | 0.715 | 0.018 | Normal | Greco et al. (2014) |
| | | | **Median** | **Mode** | | |
| $K_s$ (saturated hydraulic conductivity) | m s$^{-1}$ | [1.58E-06, 58E-06] | 9.6E-06 | 2.7E-06 | Lognormal | Pirone et al. (2016) |

$$\mu = \ln(Md), \quad \sigma = \sqrt{\ln\left(\frac{Md}{Mo}\right)} \tag{10}$$

$\gamma_d$ and $n_p$ exhibit a negative relationship. In the absence of a correlation coefficient reported in the literature, this paper assumes a correlation coefficient of 0.7. This choice addresses the strong relationship between the two variables while leaving a degree of unexpected variation due to natural fluctuations in $\gamma_d$ within the site of interest. A summary of all the pdfs describing the physical properties listed in Table 1 is illustrated by Fig.2.

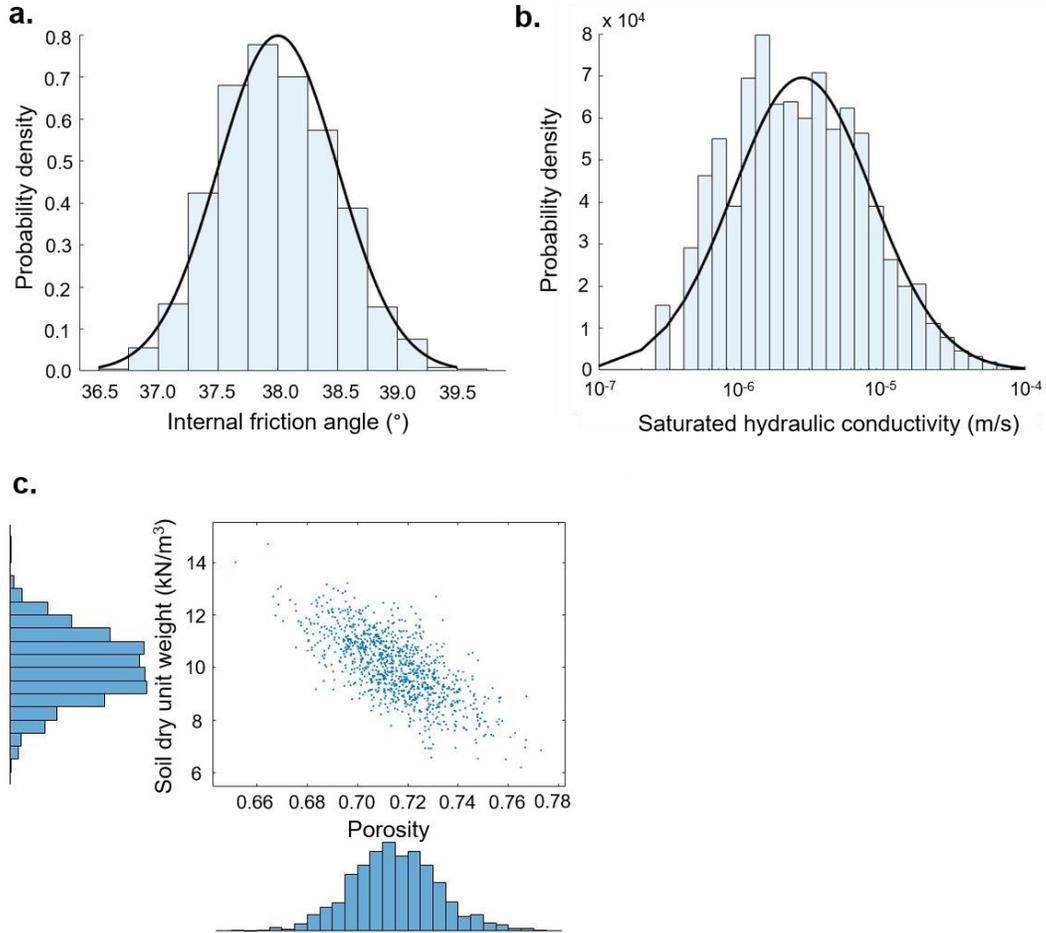

**Fig.2:** pdfs and randomized samples for Monte Carlo variables in Table 1. (a.) pdf for $\phi'$; (b.) pdf for $K_s$; (c.) $\gamma_d$ and $n_p$ correlated randomized samples;

The morphological variables included in the model are $\alpha$ and $L$. While $\alpha$ is not treated as a Monte Carlo variable, $\alpha$ and $L$ have been shown to exhibit a strong correlation (Salciarini et al., 2007; De Vita et al., 2006). From the work of De Vita et al. (2006), which focuses on the $\alpha, l$ relationship for Campanian slopes, the regression line in Fig.3a is used to model the mean soil column height ($\mu_L$) as a function of $\alpha$. The uncertainty bars, estimated to represent a variation of $0.2\mu_L$ for all $\alpha$, are used to model the standard deviation of soil column height ($\sigma_L$) where $4\sigma_L = 0.2\mu_L$. $\mu_L$, $\sigma_L$ are used as parameters to model $L$ as a normally distributed variable. Fig.3 shows the Monte Carlo generated values of $L$ superimposed to the original regression line, as well as pdfs of $L$ for different values of $\alpha$.

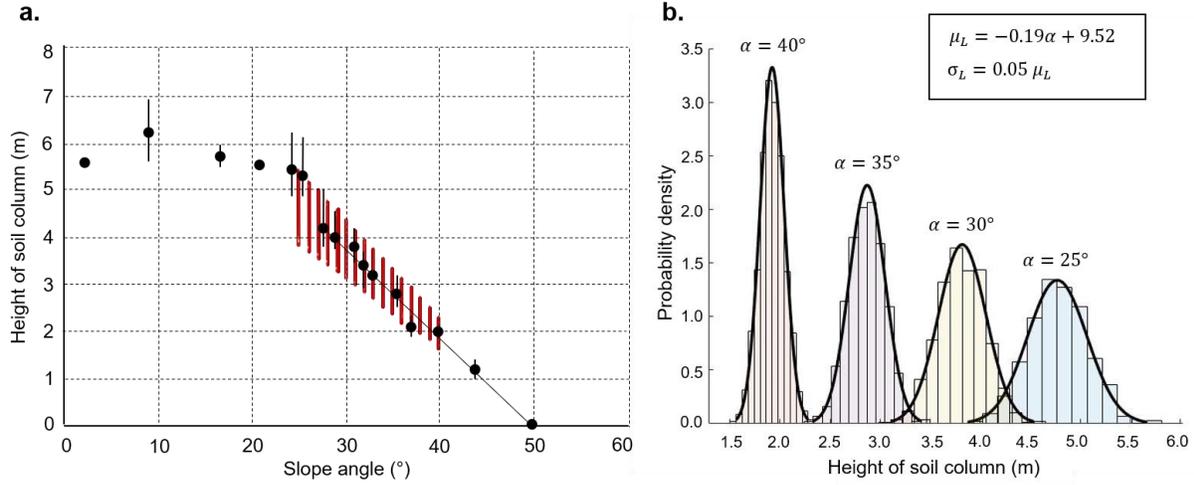

**Fig.3:** (a.) comparison of Monte Carlo-generated data points for each slope angle level (in red) to the empirical relationship between slope angle and soil column height (data from De Vita et al., 2006); (b.) pdfs of $L$ for selected values of $\alpha$.

To establish Monte Carlo initial conditions, the initial pressure head profile $h_0$ is analyzed. In the literature, a linear profile for $h_0$ is often assumed (Lizarraga et al., 2017; Gao et al., 2017; Iverson, 2000). Here, this assumption is also exploited in the context of a data-driven approach constraining $h_0$ in agreement with the typical conditions of the Campanian slopes. At this reference, the data reported by Comegna and Damiano (2016) for the Cervinara site (Fig.4a) shows a high seasonal fluctuation in pressure head, with $h$ closer to 0 (saturation) in the winter months.

The winter season was chosen as representative of the worst-case scenario for a conservative early warning approach. Although the historical database presented by Calvello and Pecoraro (2018) show that Fall, Winter and Spring months have seen similar volumes of landslide events, this occurrence is also representative of the seasonal variation in precipitation patterns in Campania, with high intensity rainfall occurring in the fall and in the spring (Rianna et al., 2018). According to this choice, the same precipitation taking place in different months is more likely to trigger landslides in the Winter, due to wetter initial condition. Still, in LEWS applications, the methodology applied for the Winter season can be repeated to construct $h_0$ pdfs for the other months; in this case, the selected season can be treated as a known external input to the model, similar to the slope angle, $\alpha$.

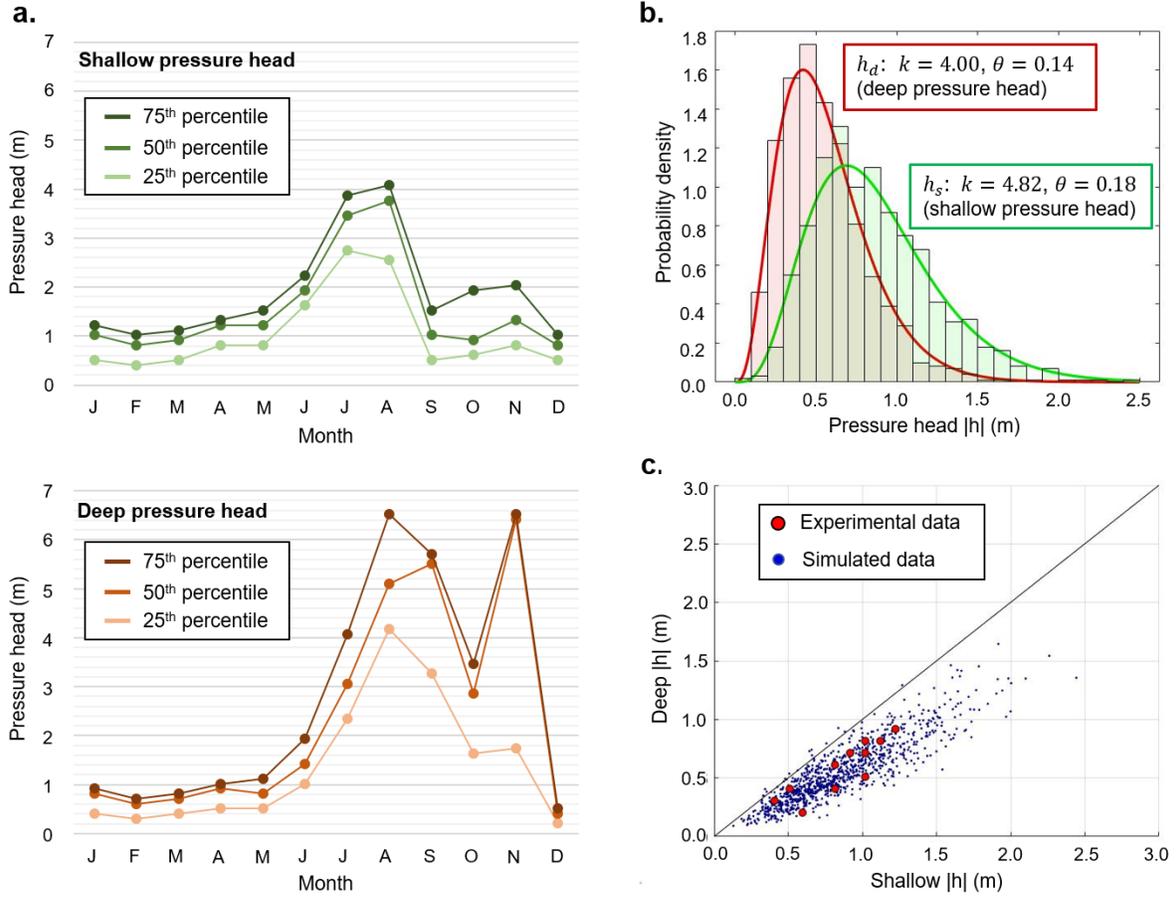

**Fig.4:** (a.) seasonal pressure head data for the Cervinara site (data from Comegna and Damiano, 2016); (b.) constructed pdfs for $h_d$ and $h_s$ (absolute value $h$ axis) and a randomized dataset; (c.) comparison of experimental and simulated $(h_d, h_s)$ pairs for pdf correlation (absolute value $h$ axes).

Pressure head data at shallow ($h_s$, $z = 0.60$ m) and deep ($h_d$, $z = L$) locations in the winter months was extracted from the database and two separate pdfs were generated (Fig.4b). Assuming that $h_0$ involves negative pore water pressure in the unsaturated zone (i.e., soil suction) and considering that the pressure data are concentrated in the 0.0-1.0 m range with a moderate distribution tail, the gamma pdf (Eq.11), defined for unsigned random variables only, was selected.

$$f(x) = \frac{1}{\Gamma(k)\theta^k} x^{k-1} e^{-\frac{x}{\theta}} \qquad (11)$$

The shape ($k$) and scale ($\theta$) parameters for $h_s$ and $h_d$ were obtained through a Monte Carlo procedure. A randomized dataset with several ($k, \theta$) pairs was generated, and the parameter pair that returned the minimum mean squared error at the percentiles in Fig.4a was selected, iteratively repeating the procedure to narrow the range of viable combinations. The pdfs obtained in Fig.4b show that at the bottom of the soil column the pressure head distribution approaches saturation. Conversely, in the near-surface, the variability in pressure head increases, possibly because of the greater influence of weather-induced variations. Fig.4b also reports the ($k, \theta$) found. Note that these values apply to a pdf of $|h| = -h$.

The data also shows a strong relationship between $h_d$ and $h_s$, with a correlation coefficient of $r \approx 0.9$ (Fig.4c). This metric was used to correlate the pdfs and generate randomized Monte Carlo pairs, compared to the experimental data in Fig.4c. Almost all $h_d, h_s$ pairs lie below the 45° line, representing that dryer initial conditions are typically found closer to the surface. Each randomized $h_d, h_s$ pair is coupled with a randomized $L$, and pressure head is linearly interpolated throughout the soil column to obtain $h_0$.

The randomized sample of $N$ data points obtained from the pdfs described in this section were subject to $n(I)$ different values of rainfall intensity combined with $n(\alpha)$ slope angles, for a total of $n(I)n(\alpha)N$ simulations. The intensities considered lie between 1 mm/h and 100 mm/h, which typically appear as the $I$-axis limits in intensity-duration plots. These values and other numerical parameters used in the model are summarized in Table 2.

**Table 2:** Summary of computational parameters used in the infiltration simulations.

| Parameter | Range / Value |
|---|---|
| $\gamma_w$ | 10 kN/m³ |
| $\theta_r$ | 0.17 (Olivares et al., 2019) |
| $b$ | 1.5 |
| $\Delta t$ | 5 s |
| $m$ | 31 nodes for each soil column |
| $I$ | Log-spaced range between 1 mm/h and 100 mm/h, $n(I) = 50$ |
| $\alpha$ | All integer slopes between 25° and 40° included, $n(\alpha) = 16$ |
| $N$ | 100 Monte Carlo combinations of randomized variables. |
| **Total simulations** | $n(I)n(\alpha)N = 80{,}000$ |

## 3. Simulation results and analysis

*3.1 – Probabilistic, slope-dependent thresholds*

Fig.5a illustrates the complete scatter plot of failure times for the simulated 80,000 infiltration events, compared to published intensity-duration thresholds for the Campania region. This large scatter cloud is visualized as a series of stacked clouds of lower variability for each simulated $\alpha$ (Fig.5b); given a $\alpha, I$ pair, the variability along the failure time axis is entirely due to Monte Carlo variables. From these visualizations, failure time variability due to both slope angle and stochastic conditions is significant and not captured by existing thresholds.

The goal of obtaining probabilistic, slope-dependent thresholds for early warning can be split into two tasks: (1) determining the mean failure time given a $\alpha, I$ pair and (2) modeling the stochastic variability around each mean as a probability density function.

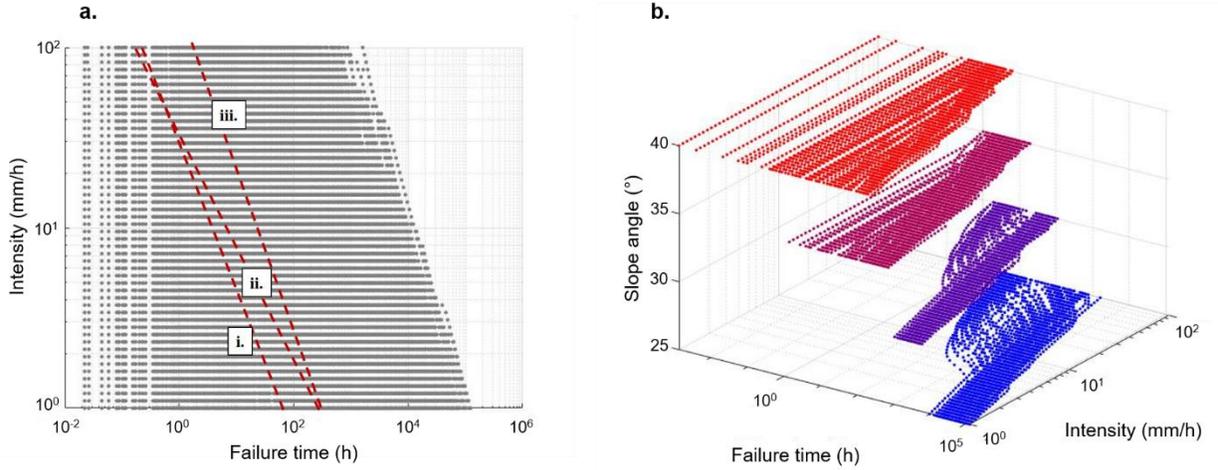

**Fig.5:** (a.) intensity vs failure time scatter plot of all simulated events and thresholds from Calcaterra et al. (2000) lower bound (i.) and upper bound (ii.), and from Guadagno (1991) (iii.); (b.) the effect of $\alpha$ on the intensity vs failure time scatter cloud for selected $\alpha$ levels (low $\alpha$ in blue, large $\alpha$ in red).

Fig.5 outlines the principal features of a $t_f$ vs $\alpha, I$ relationship:

1. The logs of $t_f$ and $I$ are linearly related, which is reasonable as total water infiltration (the product of intensity and duration) is what causes slope instability.
2. The slope of the intensity-duration average threshold is affected by $\alpha$.
3. On average, larger $\alpha$ causes a decrease in $\log t_f$ (more inclined slopes fail earlier).

This concept is formalized in Eq.12, which computes the mean log failure time $\mu_{\log t_f}$. Since $\tan \alpha$ appears in the factor of safety expression (Eq.10), the regression retains this formulation.

$$\mu_{\log t_f}(\alpha, I) = \beta_0 + \beta_1 \log I + \beta_2 \tan \alpha + \beta_3 \tan \alpha \log I \tag{12}$$

The regression leads to a relatively high explained variation ($R^2 = 0.74$) considering that the variation due to stochasticity has not been addressed. The estimated regression coefficients $\beta_0, \ldots, \beta_3$ are all statistically significant and are reported in Table 3. Fig.6 compares failure times from $\mu_{\log t_f}(\alpha, I)$ to the Monte Carlo scatter, for several $\alpha$. Qualitatively, the failure times from $\mu_{\log t_f}(\alpha, I)$ remain at the center of each $\log t_f(\alpha, I)$ cloud, making it reasonable to model such variability as a lognormal pdf with mean $\mu_{\log t_f}(\alpha, I)$. The standard deviation of failure times due to stochasticity varies in the $\alpha, I$ space (Fig.5-6). A 3D plot of the standard deviation of the logarithm of failure time $\sigma_{\log t_f}$ vs $\alpha, I$ is shown in Fig.7 together with the corresponding predicted $\sigma_{\log t_f}$ based on the regression of Eq.13.

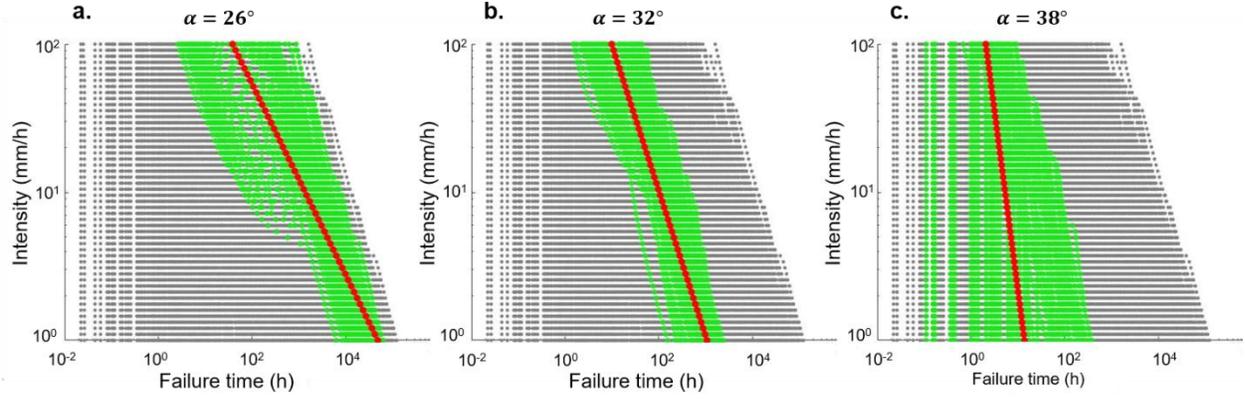

**Fig.6:** intensity vs failure time simulated data points (green) for (a.) $\alpha = 26°$, (b.) $\alpha = 32°$, (c.) $\alpha = 38°$ compared to $\mu_{\log t_f}(\alpha, I)$ prediction (red) over the entire simulated dataset (gray).

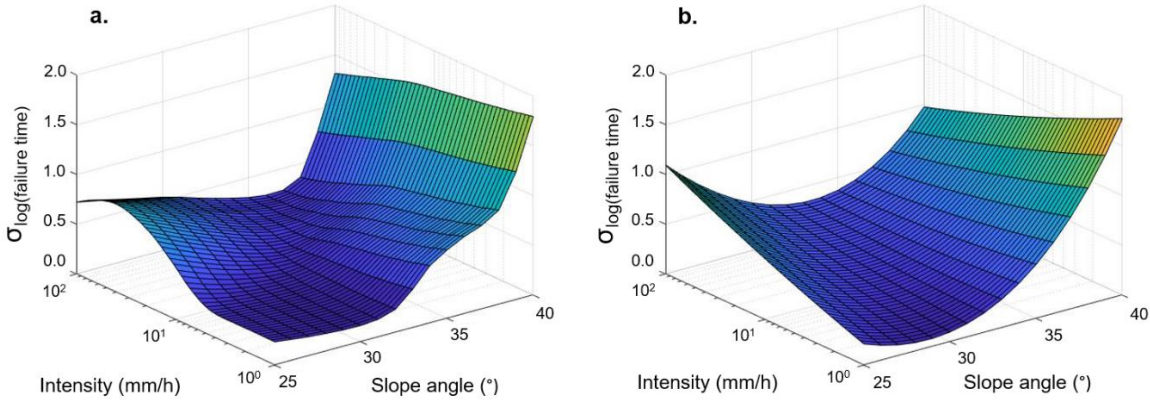

**Fig.7:** $\sigma_{\log t_f}$ vs $\alpha, I$ from (a.) simulated database, (b.) second-order approximation through regression.

A second-order regression model to approximate the experimental data (Fig.7) was constructed with the regression in Eq.13. The coefficient of determination obtained was $R^2 = 0.90$; the regression coefficients are reported in Table 3.

$$\sigma_{\log t_f}(\alpha, I) = \gamma_0 + \gamma_1 \log I + \gamma_2 (\log I)^2 + \gamma_3 \tan \alpha + \gamma_4 (\tan \alpha)^2 + \gamma_5 \tan \alpha \log I \qquad (13)$$

**Table 3:** Computed regression coefficients from Eq.12-13 and their statistical significance

| Regression coefficient | Estimated value | 95% confidence interval |
|---|---|---|
| $\beta_0$ | 10.57 | [10.52, 10.63] |
| $\beta_1$ | -3.44 | [-3.49, -3.40] |
| $\beta_2$ | -12.07 | [-12.16, -11.99] |
| $\beta_3$ | 3.88 | [3.81, 3.95] |
| $\gamma_0$ | 6.06 | [5.76, 6.37] |
| $\gamma_1$ | 1.42 | [1.32, 1.51] |
| $\gamma_2$ | 0.0389 | [0.0124, 0.0654] |
| $\gamma_3$ | -21.85 | [-22.77, -20.92] |
| $\gamma_4$ | 19.95 | [19.24, 20.65] |
| $\gamma_5$ | -2.26 | [-2.38, -2.13] |

It is important to distinguish between the model of Eq.12, which is based on mechanistic intuition and involves fewer parameters, and the model of Eq.13, which is similar to a local approximation of $\sigma_{\log t_f}(\alpha, I)$ as encountered in the simulated dataset. While Eq.12 can be reasonably exported to other case sites, Eq.13 requires further study; in this case, the main conclusion from the analysis of $\sigma_{\log t_f}(\alpha, I)$ is that the dependence on $\alpha, I$ is significant, and parametric or fully data-based models are needed to obtain this term.

Therefore, at any given slope angle and intensity, the logarithm of failure time is normally distributed:

$$\log t_f(\alpha, I) \sim \mathcal{N}\left(\mu_{\log t_f}(\alpha, I), \sigma^2_{\log t_f}(\alpha, I)\right) \tag{14}$$

The inverse problem, namely finding the probability $p$ that a real-time precipitation of intensity $I$ lasting $T$ hours triggers a failure event at slope $\alpha$, is solved by integrating the pdf (Eq.15, the subscript $\log t_f$ on the distribution parameters was dropped for clarity); this is equivalently interpreted as the probability that $t_f$ for the location and conditions considered occurs within the first $T$ hours.

$$p(T, \alpha, I) = \frac{\log e}{\sigma(\alpha, I)\sqrt{2\pi}} \int_0^T \frac{1}{t} e^{-\frac{(\log t - \mu(\alpha, I))^2}{2\sigma^2(\alpha, I)}} dt \tag{15}$$

Note that the term $\log e$ in Eq.15 is used to normalize the pdf given the use of logarithms in base 10 for intuitive visualization of intensity and failure time, versus the conventional pdf of Eq.9. Fig.8 plots the $p(T, \alpha, I)$ contours on an intensity-duration space for various $\alpha$.

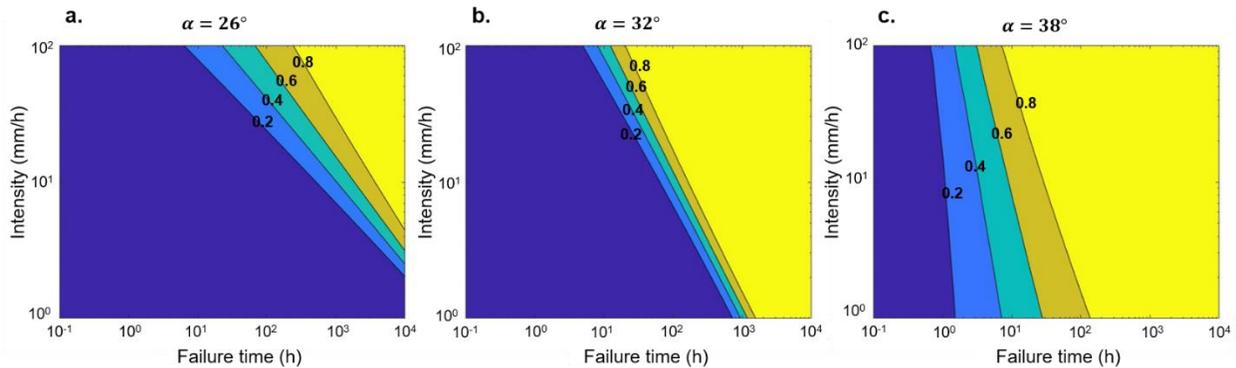

**Fig.8:** intensity-duration probabilistic thresholds for (a.) $\alpha = 26°$, (b.) $\alpha = 32°$, (c.) $\alpha = 38°$. The numbers shown correspond to the probability associated to each threshold.

The probabilistic thresholds from Fig.8 are compared across different slope angles and, for illustration purposes, overlaid to a scatter of global landslide events from Guzzetti et al. (2008) in Fig.9. It is noticeable that the threshold variability due to slope angle ($p = 0.5$ contours are plotted in red) can exceed the probabilistic variability (illustrated in different color shades in the range $0.2 \leq p \leq 0.8$) within a slope angle. At intensities of 1 mm/h, the slope angle variability causes differences in failure time of two orders of magnitude; this effect is comparable to the failure time probabilistic variability for $\alpha = 38°$, and more than one order of magnitude higher than the probabilistic scatter for $\alpha = 32°$. This result further highlights the need to differentiate landslide events and intensity-duration thresholds by slope angle for higher early warning accuracy.

The thresholds found in this work cannot be directly evaluated and compared against a landslide data scatter or against empirical thresholds, in that such step would have required use of accurate, site-specific measurements of slope angle, not available to the authors. Nevertheless, the single-angle thresholds obtained from the simulations display a weaker dependency on rainfall intensity for landslide triggering than the empirical thresholds from the literature. A possible explanation for this observation is that the gradient of the empirical thresholds may be lowered by the inclusion of different slope angles that stretch the data horizontally (Fig.9). In future efforts to construct physics-based early warning systems, more accurate data reporting to account for slope angle will be critical in analyzing this trend.

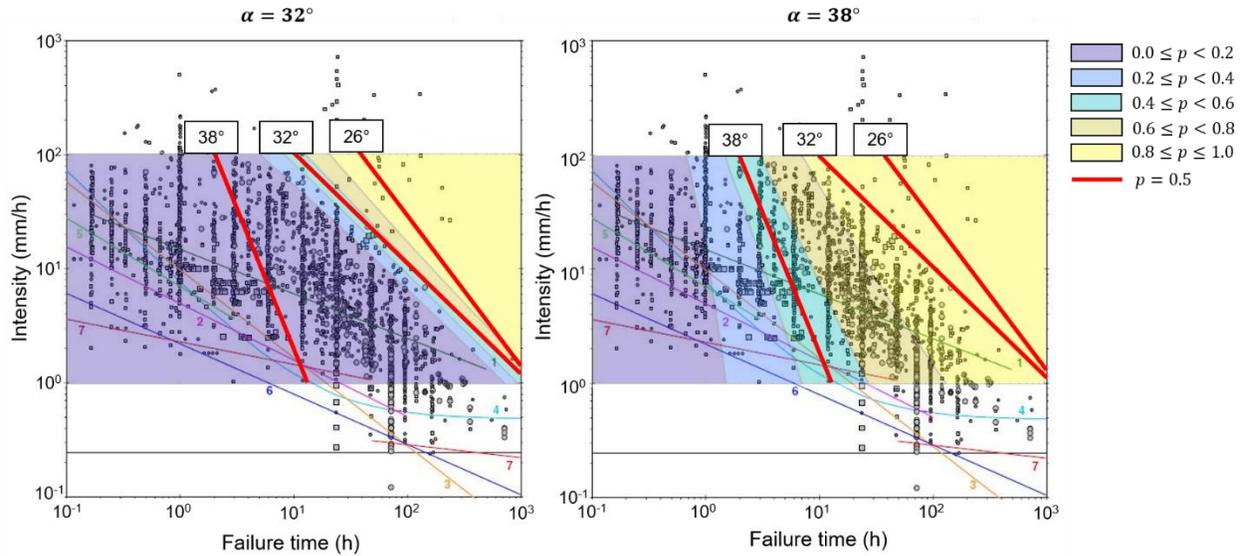

**Fig.9:** in color shades, probabilistic thresholds for shallow landslide initiation for $\alpha = 32°$ (left) and $\alpha = 38°$ (right) overlaid on global landslide events data scatter from Guzzetti et al. (2008), showing the large real-world variability of failure time events. To compare the stochastic scatter to the slope angle-induced failure time variability, the $p = 0.5$ probability thresholds for $\alpha = 26°, 32°, 38°$ are plotted in red.

*3.2 – Relative variable importance*

For accurate real-time decision making, a lower variability between probabilistic thresholds can give better confidence on when early warning alerts should be issued: Fig.8b would be more effective than Fig.8c at this task as the probabilistic thresholds in the latter span various orders of magnitude. Ultimately, the applicability of the model depends on $\sigma_{\log t_f}(\alpha, I)$: if the standard deviation is too high, $p(T, \alpha, I)$ approaches similar values throughout the intensity-duration space considered. On the other hand, if we theoretically monitor a single point in space for which all variables are perfectly known (no stochasticity), $\sigma_{\log t_f}(\alpha, I)$ becomes zero and $p(T, \alpha, I)$ forms a binary threshold. While achieving the latter case is unfeasible, reducing $\sigma_{\log t_f}(\alpha, I)$ can be done by decreasing the variance of Monte Carlo inputs from Fig.2-4. The spatial variability of the stochastic variables can be reduced by dividing a monitored region into several sub-regions with different pdfs; the variability due to uncertainty can be reduced by conducting more detailed studies on the region. Considering these possible improvements, this section shows how the Monte Carlo inputs to the model can be ranked in terms of their impact on failure time variability so that, under time or budget constraints, a user interested in constructing intensity-duration thresholds using the methodology of section 3.1 for any location of interest can focus their efforts on primarily studying one or few variables.

The algorithm used to rank aleatory variables consists of the following steps: (1) dimensionality reduction through Principal Component Analysis (PCA) of the set of stochastic variables in case of multicollinearity; (2) normalization of the reduced set of variables through z-scores; (3) regression of $\log t_f$ against the normalized stochastic variables at each $\alpha, I$ level; (4) ranking variables at each $\alpha, I$ level according to the magnitude of the corresponding regression coefficients. With this algorithm, the importance of each factor varies over the $\alpha, I$ space: depending on the typical meteorological and topographic characteristics of the region of interest, different aleatory variables can emerge as most critical. The normalization step ensures that the regression coefficients do not depend on units or on their pdf variance.

In this application, PCA is applied to transform highly correlated $h_d$, $h_s$ data into a single array data of initial conditions, $\widehat{h_{PC}}$. The same procedure can be generally applied to highly correlated variables before performing the regression step; here, $\gamma_d$ and $n_p$ were not reduced through PCA as their correlation was not high enough to justify an intrinsically one-dimensional system. Following reduction and normalization, at each $\alpha, I$ the regression on the normalized stochastic variables is performed (Eq.16). Note that this regression is not locally accurate at predicting failure times, as doing so would require a more sophisticated expression or a neural network approach, which may result in overfitting; however, from a first-order analysis it is possible to estimate which variables have the most significant impact.

$$\log t_f (\alpha, I) = \delta_0 + \delta_1 \widehat{\phi'} + \delta_2 \widehat{n_p} + \delta_3 \widehat{\gamma_d} + \delta_4 \widehat{K_s} + \delta_5 \widehat{L} + \delta_6 \widehat{h_{PC}} \qquad (16)$$

The magnitudes of the regression weights can be compared to each other (by normalizing their sum at each $\alpha, I$ pair to 1) and provide a measure of the relative importance of each variable. Fig.10a illustrates, over the $\alpha, I$ space, the linear combination of the regression weight and the color assigned to each variable per the legend in the figure. This illustration is qualitative and demonstrates the presence of different regimes of variable importance in the range of slope angles and rainfall intensities analyzed. In order to show the effect of each factor, Fig.10b is a more convenient visualization as it captures the changes in the contribution of the aleatory variables analyzed within each regime.

Fig.10 shows that specific dry weight, saturated hydraulic conductivity and initial suction conditions are the aleatory variables that most affect the variability of failure times. $K_s$ is shown to be most significant (green) when the failure events are characterized by high infiltration (high $I$) and lower dependence on initial conditions (low $\alpha$ delays failure events), as low hydraulic conductivity can limit and delay the amount of water sinking in the column. For failure events happening at large failure times (low inclination and low infiltration), the significance of $\gamma_d$ is highest (yellow); considering Eq.6, this result can be interpreted as the increase of stress due to weight dominating over the strength decay due to suction reduction. On the other hand, when $\alpha$ is high, small changes in pressure head can lead to slope instability much quicker, making the system more reliant on initial conditions rather than on rainfall (purple). The transition between the $K_s/\gamma_d$ and $h_{PC}$ regimes occurs as the slope angle approaches the internal friction angle and the system becomes suction-stabilized in the unsaturated domain.

Given the correlation between dry weight and porosity (Fig.2c), in the $\gamma_d$ regime $n_p$ can also be labeled as a significant variable; still, the regression output suggests that independent changes in $\gamma_d$ have a larger impact on failure times than variations in $n_p$. A choice of correlation coefficient between $n_p$ and $\gamma_d$ closer to 1 could turn this into a one-dimensional system where a PCA-reduced variable is used in the regression step and both factors are reported as equally significant in their regime. The findings reported in this section can guide research strategies in potential sites of application of LEWS: in tropical areas characterized by intense storms, knowing $K_s$ with high accuracy brings important advantages to the ability of predicting failure events; instead, for locations where precipitations are less intense and slopes are shallow, studying $\gamma_d$ could be more relevant.

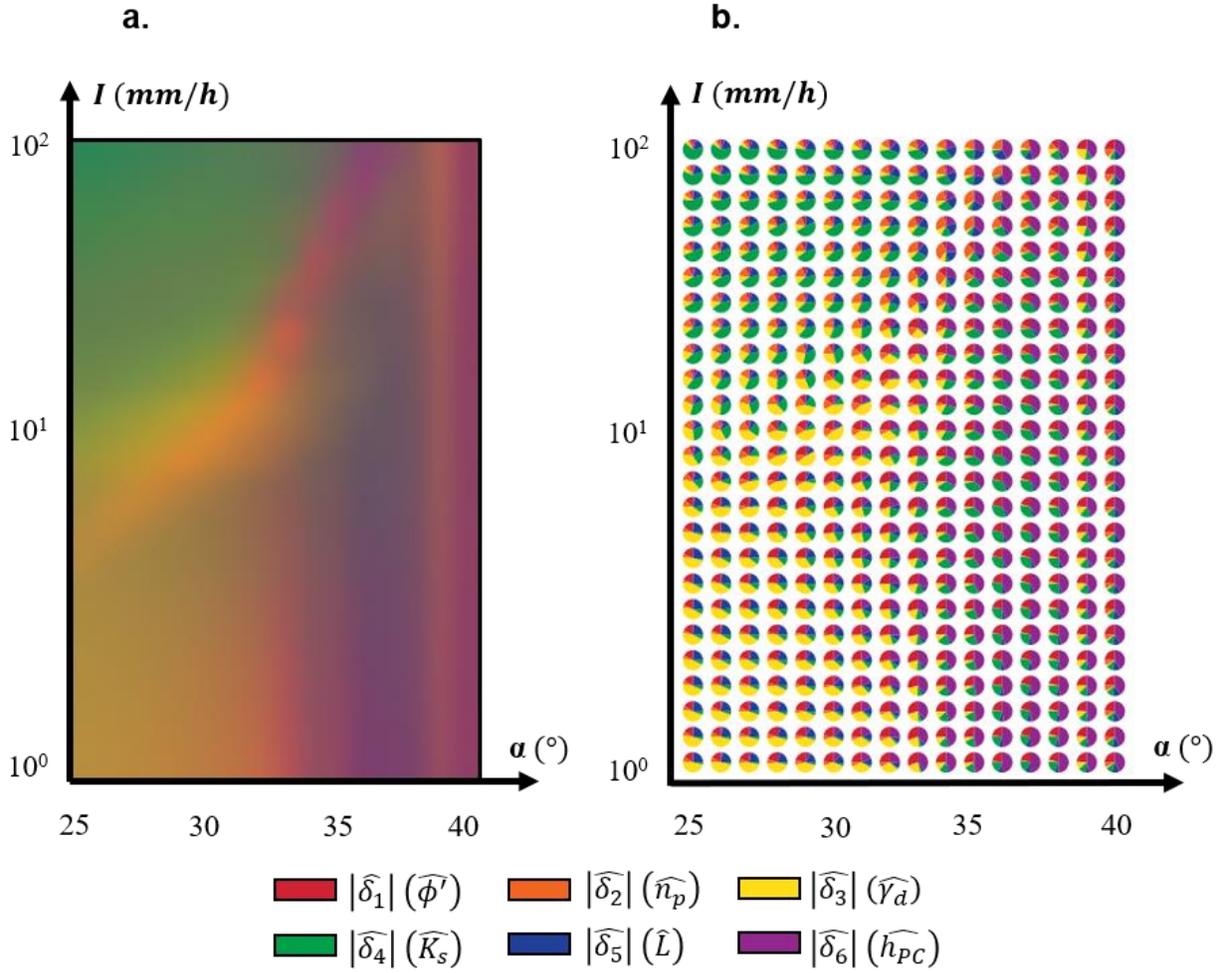

**Fig.10:** (a.) linear combination of regression weights and the corresponding colors associated to its entries over the $\alpha, I$ space, showing the three main regimes of variable importance ($K_s, \gamma_d, h_{PC}$); (b). pie charts of the regression weights over the $\alpha, I$ space. $|\hat{\delta}_i|$ denotes the normalized magnitude of the regression weight $\delta_i$.

## 4. Conclusions

By blending physical concepts embedded into a hydro-mechanical model and a rigorous stochastic data treatment, a physically-based, yet probabilistic, assessment of the risk of shallow landslide initiation was conducted. The approach proposed is user-centered: failure time thresholds explicitly incorporate information that is readily available while monitoring a site (rainfall intensity and slope angle), while quantities that are known with higher uncertainty and involve a more complex data collection setup (initial suction conditions; mechanical and hydraulic soil parameters) are embedded as probabilistic variables. After modeling the probability density functions of each stochastic variable using available data, as done for the case of the Campanian soils, deterministic and probabilistic inputs were provided to a computational tool simulating water infiltration in one-dimensional soil columns until failure.

The simulation output (log of failure time) at each slope angle-intensity pair was modeled as a normally distributed random variable, with mean and standard deviation predicted by a global regression. The probability density function of $\log t_f$ was integrated over rainfall duration to calculate the real-time probability of landslide triggering at given slope angle and rainfall intensity. When overlaying probabilistic thresholds for one slope angle to single thresholds for various angles, the failure time variability due to slope angle appeared to be as significant as the variability introduced by all other Monte Carlo inputs together. This should inspire systematic reporting of slope angle in landslide databases so that empirical slope-angle dependent thresholds can be constructed and compared to the existing computational simulations, such as the one developed for this paper.

This synthetic, stochastic model can be applied to monitor in real-time, precursors of landslide triggering over a landscape of different slopes. For higher confidence, the probabilistic variability should be reduced through the construction of more accurate and/or segmented probability density functions for the stochastic variables over the region of interest. An algorithm to rank the impact of each Monte Carlo variable on the probabilistic variability was implemented, regressing $\log t_f$ over a set of reduced, normalized Monte Carlo variables at each $\alpha, I$, and ranking variables in terms of the magnitude of the corresponding regression coefficients.

This approach showed that soil specific dry weight, saturated hydraulic conductivity and initial suction conditions had the most significant impact on the probabilistic variability. Hydraulic conductivity was most relevant for high intensity and low slope angle combinations, where slope instability occurs relatively late after significant infiltration; initial suction had higher relative importance on inclined slopes, where failures occur earlier at a low cumulative water infiltration and are thus more dependent on initial conditions; dry weight, correlated to porosity, is most important for events with low infiltrations and large failure times, for stress increase through the soil column weight is larger than the strength decrease due to suction reduction. This idea can help researchers prioritize studies on specific mechanical, hydraulic and initial conditions variables depending on the known typical rainfall characteristics and slope angle distribution for a site of interest.

The conclusions drawn in this paper on the regression form of the average $\log t_f$ and on which stochastic variables are most significant are mechanistically justified and emerge from models that avoid overfitting to the case study of the Campanian soils. Future research would involve performing a sensitivity analysis of this methodology by applying it to other sites of interest for which the literature provides data. Most importantly, as more sophisticated infiltration models and factor of safety evaluations emerge, the method presented in this paper can further improve in the fidelity of how the simulated data represents real-world landslide triggering phenomena.

## Declarations

**Funding**

This work was supported by Grant No. ICER-1854951 awarded by the U.S. National Science Foundation.

**Competing interests**

The authors declare that they have no known competing financial interests or personal relationships that could have appeared to influence the work reported in this paper.

**Availability of data and material**

Not applicable

**Code availability**

Not applicable

**Appendix A: Equations embedded in the water infiltration solver**

Eq.A1-A4 show how each updated pressure head profile is calculated following a time step $\Delta t$:

$$\underline{h_{t+1}} = \begin{bmatrix} h_{1,t+1} \\ \vdots \\ h_{i,t+1} \\ \vdots \\ h_{m,t+1} \end{bmatrix} \quad (A1)$$

where:

$$h_{i,t+1} = \frac{\left[\overline{(K_{\iota,\iota-1})}_t - \overline{(K_{\iota,\iota+1})}_t - \frac{1}{\Delta z}\overline{(K_{\iota,\iota-1})}_t(h_{i,t} - h_{i-1,t}) - \frac{1}{\Delta z}\overline{(K_{\iota,\iota+1})}_t(h_{i,t} - h_{i+1,t})\right]\Delta t}{\left[\overline{(C_{w_{\iota,\iota-1}})}_t + \overline{(C_{w_{\iota,\iota+1}})}_t\right]\frac{\Delta z}{2}} + h_{i,t} \quad (A2)$$

$$h_{1,t+1} = \frac{\left[I \cos\alpha - \overline{(K_{1,2})}_t - \frac{1}{\Delta z}\overline{(K_{1,2})}_t(h_{1,t} - h_{2,t})\right]\Delta t}{\overline{(C_{w_{1,2}})}_t \frac{\Delta z}{2}} + h_{1,t} \quad (A3)$$

$$h_{m,t+1} = \frac{\left[\overline{(K_{m,m-1})}_t - \frac{1}{\Delta z}\overline{(K_{m,m-1})}_t(h_{m,t} - h_{m-1,t})\right]\Delta t}{\overline{(C_{w_{m,m-1}})}_t \frac{\Delta z}{2}} + h_{m,t} \quad (A4)$$

where Eq.A5 defines the following operation for a generic nodal variable $x$:

$$\overline{(x_{\iota,J})}_t = \frac{x_{i,t} + x_{j,t}}{2} \quad (A5)$$

Similarly, $FS(z,t)$ is also discretized to be evaluated at each node $i$ (Eq.A6-A7):

$$FS_{i,t} = \frac{\tan\phi'}{\tan\alpha}\left(1 - \frac{\gamma_w h_{i,t} S_{r_{i,t}}}{\sigma_{i,t}^{net}}\right) \quad (A6)$$

$$\sigma_{i,t}^{net} = z_i \gamma_d + n_p \gamma_w \Delta z \sum_{j=2}^{i}\overline{(S_{r_{J-1,J}})}_t \quad (A7)$$